# Moiré and beyond in transition metal dichalcogenide twisted bilayers

Kha Tran[1,*], Junho Choi[2], Akshay Singh[3,*]

NRC research associate at the Naval Research Laboratory, 4555 Overlook Ave SW, Washington, DC 20375, USA

Department of Physics, The University of Texas at Austin, Austin, TX 78712, USA

Department of Physics, Indian Institute of Science, Bengaluru, Karnataka -560012, India

*-corresponding authors: kha.tran.ctr@nrl.navy.mil, aksy@iisc.ac.in

Abstract: Fabricating van der Waals (vdW) bilayer heterostructures (BL-HS) by stacking the same or different two-dimensional (2D) layers, offers a unique physical system with rich electronic and optical properties. Twist-angle between component layers has emerged as a remarkable parameter that can control the period of lateral confinement, and nature of the exciton (Coulomb bound electron-hole pair) in reciprocal space thus creating exotic physical states including moiré excitons. In this review article, we focus on opto-electronic properties of excitons in transition metal dichalcogenide (TMD) semiconductor twisted BL-HS. We look at existing evidence of moiré excitons in localized and strongly correlated states, and at nanoscale mapping of moiré superlattice and lattice-reconstruction. This review will be helpful in guiding the community as well as motivating work in areas such as near-field optical measurements and controlling the creation of novel physical states.

## Introduction

The scientific community has witnessed tremendous work in the semiconducting 2D-TMD layered materials space in the past 10 years. In the spirit of the facile equation 2D+2D=2D, significant progress has been made in fabricating vdW BL-HS by stacking the same or different component layers[1-3]. Such a BL-HS has many degrees of freedom including layer composition, physical nature (metallic, semiconducting), method of preparation, twist-angle and stacking order. Through a combination of these degrees of freedom, new applications towards quantum information, opto-electronic transport and integrated-photonics are expected [4-7].

Quantum information has advanced from a strictly-academic to an industrially-relevant endeavor, with recent work claiming quantum supremacy in the case of certain algorithms[8]. However, to create scalable computing and communication systems, there is a critical need of reliable, environmentally-protected and deterministically-located solid-state single photon sources. In parallel, there is an increased focus on integration of advanced 2D nanomaterials in traditional electronic circuits, and to understand and control electronic and thermal transport. For example, by interfacing various 2D materials, ultra-high thermal insulation has been observed, which can be exploited for thermal management and energy harvesting[9].

Twisted BL-HS offers novel control over opto-electronic properties, via control of periodicity and correlations. Analogous to twisting two polarizer sheets to control transmission of light, twisted BL-HS can control optical and electronic transport. Additionally, like taking a picture of a pixelated screen, an emergent moiré structure with beautiful patterns and high symmetry points offers a new super-structure. In practice, for BL-HS, tunable twist-angle leads to modification of reciprocal space , momentum mismatch, change of interlayer spacing, Coulomb interactions and tunable correlations.

Moiré superlattices can form when two MLs having slightly different lattice constant and/or finite twist-angle are stacked vertically. Figure 1 center panel shows a schematic of the moiré superlattice in BL-HS with different twist angles. The unit cell of moiré superlattice is the moiré supercell. The size of the supercell $a_M$, as a function of twist-angle $\theta$ is given by[10]

$$a_M = \frac{(1+\delta)a_0}{\sqrt{2(1+\delta)(1-cos\theta)+\delta^2}}$$

, where $\delta$ is the lattice mismatch defined as $|a_0' - a_0|/a_0$; $a_0$ and $a_0'$ are the lattice constants of the two constituent MLs ($a_0 < a_0'$). For near-zero twist-angle in a small lattice mismatch BL-HS, size of the moiré supercell can be quite large i.e. tens of nanometers as compared to Bohr radius of the exciton (few nanometers). Thus, the exciton can be described as a particle moving in a slowly varying moiré potential[11,12].

There are several questions that the 2D materials twistronics community has endeavored to answer in the recent years. How are optical properties of an exciton modified in the presence of another layer, beyond a simple dielectric environment change? Can we have reliable single photon emitters in 2D materials? Can component layers be considered as separate layers, or as a vdW solid? Can spin-valley coupling result in novel physics in BL-HS? We have strived to examine possible answers and have organized this review to cover fundamental and application-relevant concepts in twisted TMD BL-HS. We briefly discuss HS preparation, especially focusing on controlling and measuring twist. Figure 1 lays out the scheme in which our review is structured. We review properties of interlayer excitons (IELXs), intralayer excitons (IALXs), and hybridized excitons (HXs) modulated by the moiré potential. We examine the effect of moiré superlattice on exciton diffusion, as well as discuss possibilities of investigating Hubbard model physics and measuring strongly correlated states. We finally review the possibility of reconstruction in small twist-angle BL-HS, which goes beyond the rigid moiré superlattice picture.

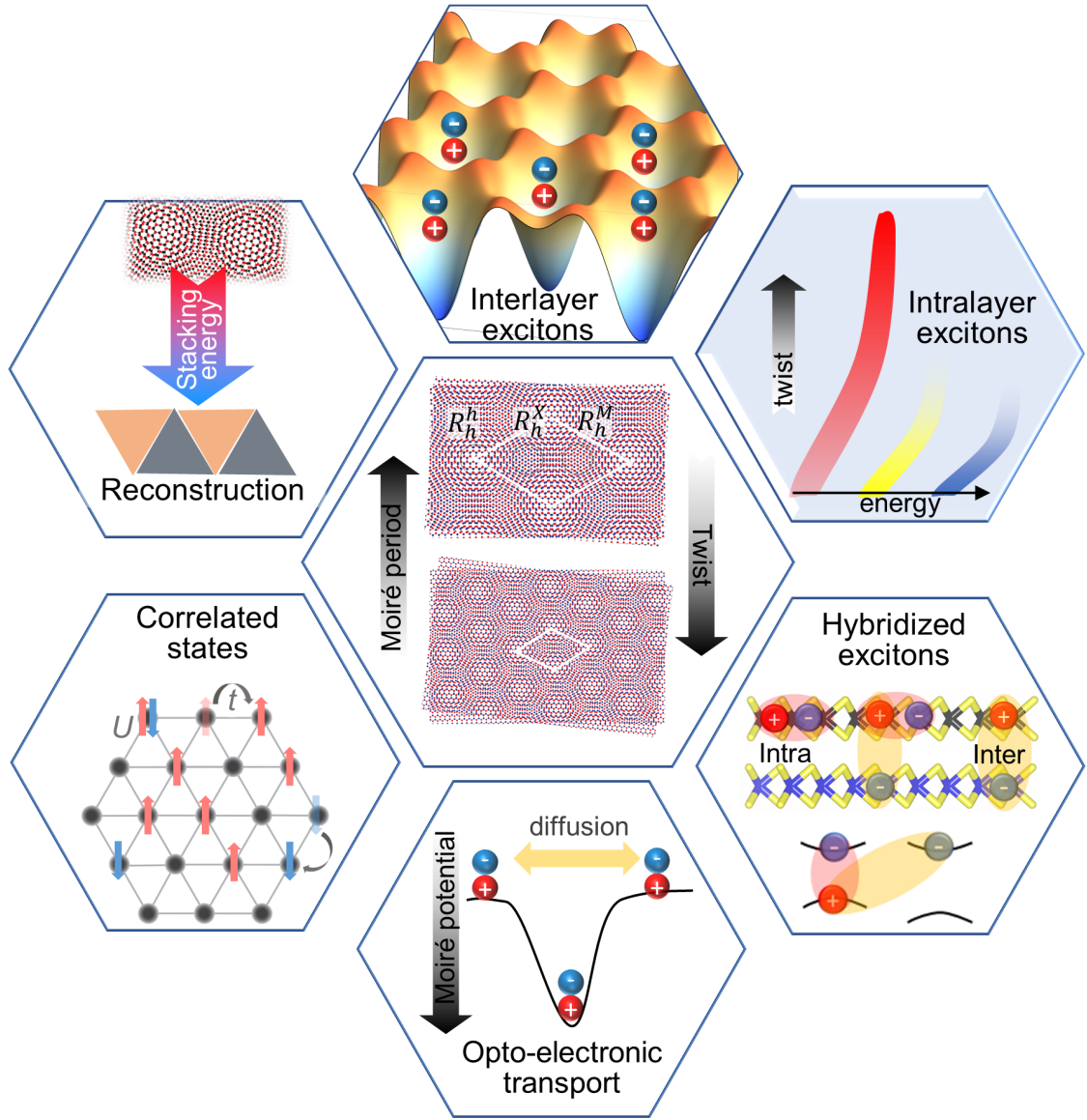

**Figure 1.** Schematic illustration of opto-electronic and structural properties of bilayer heterostructures (BL-HS) modulated by moiré superlattice and beyond. Presence of moiré superlattice in semiconducting BL-HS introduces new quantum phenomena for interlayer excitons (IELX), intralayer excitons (IALX), and hybridized excitons (HXs) controlled by twist-angle. Strongly correlated states in moiré superlattice, and stacking-energy reduction by formation of reconstructed domains opens new avenues for novel physical states.

**Twisted bilayer-heterostructure preparation**

Since a number of review papers have already discussed the basics of TMD BL-HS preparation, we will not strive to cover all aspects of HS preparation[1,3]. We only focus on methods to control the twist (azimuthal angle) between component monolayers (MLs) of a BL-HS. We note that sequential preparation of a twisted BL-HS using chemical vapor deposition (CVD) is not possible for small twist-angles < 2-5°. This is because the layers conformally grow, and can undergo lattice expansion (or contraction) for epitaxial or thermodynamically favorable-orientation growth[13].

Therefore, the dominant and most suitable method for twist-controlled BL-HS preparation is sequential mechanical stacking of exfoliated or CVD-synthesized MLs using dry[14,15] or wet transfer methods.[1,3,16,17] Since encapsulation by hexagonal-boron nitride (hBN) improves opto-electronic properties, including narrowing of linewidths and reducing surface charge, we will introduce how to fabricate an hBN encapsulated BL-HS by the dry transfer technique[14,18]. Briefly, a polymer mask (PDMS or PPC) contacts the top hBN layer (exfoliated on a silicon or polymer substrate) at an elevated temperature. The top hBN layer is picked up, then TMD ML1, then ML2 and finally the bottom hBN. The encapsulated BL-HS drops onto a substrate upon contact and heating (to temperature beyond the polymer softening temperature). A solvent step follows to remove any polymer residue, followed by annealing in vacuum or inert conditions.

Coarse control of twist-angle during stacking is provided via straight-edge of CVD ML or fracture/tear along a crystal orientation (zigzag or armchair) for mechanically exfoliated ML. Orientation of the constituent MLs is determined before stacking by measuring second harmonic generation, and twist-angle alignment is performed accordingly. In this way, the relative twist can be controlled with an accuracy of ~ 0.5 -1° [19,20]. Geometric guidance can serve to improve the angle accuracy to better than 1° by using larger crystals[21]. This technique was recently extended to centimeter scale HS[22]. For preparing homo BL-HS, mechanical tearing of part of the ML, rotation, and subsequent stacking back (on the parent ML) can be applied for high accuracy ~ 0.2° [15,23]. An alternative is preparing an upper rotatable layer, and rotation with a microscopic tip[10].

**Opto-electronic properties of moiré excitons in twisted TMD BL-HS**

Throughout this review, we denote the parallel (antiparallel) stacking -near 0° (60°) twist-angle as R-type (H-type). Within the moiré supercell, there are three locations $R_h^h$, $R_h^X$, and $R_h^M$ ($H_h^h$, $H_h^X$, and $H_h^M$) in which the three-fold rotational symmetry is preserved (figure 1 center panel). $R_h^\mu$ ($H_h^\mu$) refers to $R$-type (H-type) stacking with the $\mu$ site (h – hexagonal center, X – chalcogen atom, M – metal atom) of the top layer aligning with the hexagonal center ($h$) of the bottom layer[11,24].

Further, due to different band gap and the work function of two constituent MLs, TMD BL-HS can have type-II band alignment[25-27]. Both IALX (electron and hole in the same layer) and IELX (electron and hole in different layers) can be present. IELX has much weaker optical dipole moment than that of IALX and, has much longer lifetime due to the spatial separation of electron and hole[26,28-30]. In the simplest case with zero twist-angle, the bandgap is direct, giving rise to direct momentum IELX. As the twist-angle increases, the conduction band (CB) minimum moves away from the valence band (VB) maximum at K point[31]. The IELX becomes indirect in momentum space with decreasing dipole moment, decreasing emission intensity[32,33] and increasing the lifetime[34].

As we will see in the following sections, moiré superlattice changes the optical and electronic properties of IELX and IALX drastically. It can localize excitons and lead to spatial variation of optical selection rules in IELXs[11,20,35,36], form topological exciton bands leading to additional satellite IALX peaks[19,37], and form triangular lattice Hubbard models for electrons or holes that simulate strong-correlation physics[38-41].

**Interlayer Moiré excitons**

Moiré superlattice imposes a periodic potential on the IELX. In a study of a rotationally aligned $MoS_2$-$WSe_2$ BL-HS[42], the local bandgap spatial variation was directly correlated with the local atomic alignment of the two MLs. The bandgap is modulated with a potential depth of ~150 meV in the lateral direction. The local maxima and minima of bandgap coincide with high symmetry sites of the moiré superlattice (figure 2(a)). Therefore, multiple IELX resonances are expected to form due to the moiré lateral confinement[43,44]. The moiré potential is strongest (~150 meV) near 0° twist-angle (R-type), and is estimated to be much weaker ~10-20 meV near 60° (H-type) [12,35,45]. 2D array of identical quantum emitters is expected to form in a deep moiré potential with a long moiré superlattice period[11].

Unlike the IALX, both the spin-conserved (singlet) and spin-flip (triplet) IELXs can couple to in-plane and out-of-plane polarized light (figure 2(h,i)) due to breaking of the out-of-plane mirror symmetry[46]. This relaxation of spin selection rule described by Yu et. al. consequently results in comparable transition dipole moments for singlet and triplet IELXs, which can then be equally bright in an optical measurement.

Moiré superlattice imposes a spatial variation of optical selection rules on both spin-singlet and spin-triplet IELX. The optical dipole moment can point either in-plane ($\sigma_+/\sigma_-$ polarizations) or out-of-plane (z polarization) depending on where the IELX is located within the moiré superlattice[24]. Only at the three-fold symmetry sites, the IELX is fully circular or out-of-plane polarized; selection rules are summarized in figure 2(i) for 0° and 60° twist-angles. In between these sites, the selection rules vary continuously within the moiré supercell[24]. Figure 2(b) maps spatial variation of the in-plane degree of circular polarization $(\sigma_+ - \sigma_-)/(\sigma_+ + \sigma_-)$, for the spin-singlet IELX. The selection rules of interlayer moiré excitons (MXs) are few of many important clues for understanding the physical origin of MXs, as we see below.

The main difficulty of finding direct evidence for interlayer MXs in optical experiments is due to limited spatial resolution. Within a laser spot size (~1 μm), there are thousands of moiré supercells, so the effect of inhomogeneous broadening (due to strain and impurities) on measured optical properties is unavoidable. Nevertheless, various research groups have relied on other clues (for example: spectral resolution, circular polarization (CP), g-factor) to support their claim of interlayer MX observation. Although studying the similar HS, mainly hBN encapsulated $MoSe_2$-$WSe_2$ HS, the optical results and interpretations vary vastly across different research groups. Below, we attempt to classify these studies based on the similarity of their main results and interpretations.

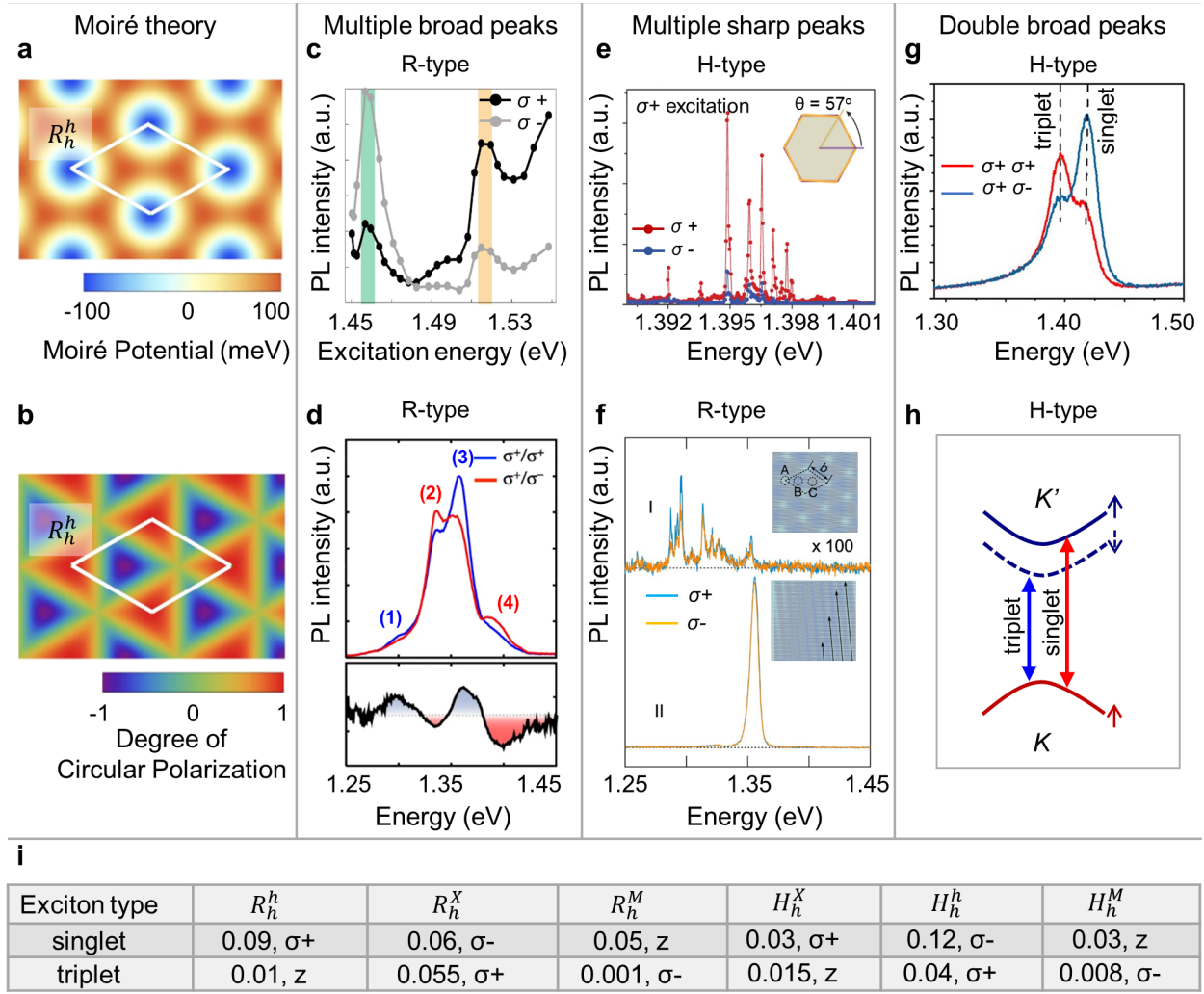

**Figure 2**. Interlayer moiré exciton (MX) theory and observation in TMD BL-HS. **a** Theoretical calculation of moiré potential (in real space) for interlayer exciton (IELX). **b** Theoretical calculation of spatial variation of the optical selection rule for in-plane polarized singlet IELX. **c,** Photoluminescence excitation (PLE) spectrum of a $WS_2$-$WSe_2$ BL-HS showing multiple resonances with opposite circular polarization (CP). **d-g,** PL spectra of $MoSe_2$-$WSe_2$ BL-HS classified by different interpretations. **h,** Schematic showing singlet (spin-conserved) transition, and triplet (spin-flipped) transition **i,** Singlet and triplet IELX dipole moment magnitude and orientation at different high symmetry sites within the moiré superlattice, for parallel stacking (R type) and anti-parallel stacking (H type)[24]. Panels

reprinted and adapted with permission, **a-b**,**d** from ref[20] (Springer Nature), **c** from ref[47] (Springer Nature), **e** from ref[36] (Springer Nature), **f** from ref[48] (Springer Nature), **g** from ref[49] (American Chemical Society).

*Broad peaks with alternating/opposite circular polarization*

Tran et al[20] have reported multiple IELX resonances in photoluminescence (PL) measurements of hBN encapsulated MoSe$_2$-WSe$_2$ BL-HS at low temperature. These peaks are attributed to the ground state and excited states of singlet IELXs, all confined in the moiré potential at site $R_h^h$. Analogous to the hydrogen atom, ground state (lowest energy peak) IELX has s-like wavefunction. The next higher energy resonance has p-like wavefunction, and so on (d-like, f-like). The IELXs inherit the valley selection rule based on the spatial distribution of their wavefunctions within the moiré superlattice[11]. The peak (1) with lowest energy has s-wave symmetry centered at the $R_h^h$ site, and its optical matrix element has $\sigma_+$ component only. The peak (2) has chiral p-wave form with an additional angular momentum, and thus mainly has $\sigma_-$ component. Similar behaviors can be inferred for peaks (3) and (4) [35]. Consequently, these peaks exhibit alternating selection rules upon PL emission (figure 2(d)). The attribution of interlayer MX to spin-triplet interlayer resonances are ruled out based on the temperature dependence and PL lifetime measurements.

Jin et al [47] have reported multiple IELX resonances in hBN encapsulated WS$_2$-WSe$_2$ BL-HS at near zero twist-angle. Although PL measurements only reveal one IELX peak at 1.43 eV, PL excitation (PLE, figure 2(c)) and resonant pump-probe measurements show the emergence of two higher energy IELX peaks (1.46 eV and 1.51 eV) with opposite CP. Their oscillator strengths are revealed directly by the pump-probe measurement and are 100 times smaller than the IALX transitions. Due to the energy separation, CP, and relative oscillator strengths among IELX peaks, 1.43 eV and 1.46 eV peaks are attributed to spin-triplet and spin-singlet transitions localized at $R_h^X$ site respectively; 1.51 eV peak is attributed to the spin-singlet transition at $R_h^h$ site. Spin-triplet transition at site $R_h^h$ is not observed.

*Sharp peaks with uniform circular polarization*

Numerous studies have reported moiré potential trapped IELXs with very sharp linewidth ~100 µeV measured by PL at low excitation power (tens of nanowatts)[36,48,50-52]. The resonances span approximately 100 meV, and are attributed to excitons trapped in the moiré potential because of their sharp linewidth and their uniform g-factor, -16 for anti-parallel stacking (7 for parallel stacking) which is similar to the inhomogeneously broadened free IELX broad peak[36,50,51]. Interestingly, under circularly polarized excitation, all the peaks exhibit uniform CP either all negative or all positive, depending on the twist-angle, suggesting similar physical origin within the moiré superlattice[36] (figure 2(e-f)). Baek et. al.[50] provides evidence of photon antibunching nature of the moiré trapped IELX in an anti-parallel stacked $MoSe_2$-$WSe_2$ BL-HS by measuring coincidence value $g^2(0)$ below 0.5. Moreover, an out-of-plane electric field can tune these quantum emitters up to 40 meV[50], suggesting the possibility of resonant cavity physics. Further, Li et. al.[52] report the possibility of interlayer biexciton (IELXX), supported by super-linear excitation power dependence of PL intensity. At the same location within the moiré superlattice, the IELXX energy is 1 to 5 meV higher than that of the IELX due to the dipolar repulsive interaction. The variation in interaction strength is attributed to the variation in confinement lengths and inter-excitonic distances of the localized IELX[52].

Bai[48] et. al. relates PL signals of MXs to real space imaging of moiré superlattice revealed by piezo force microscopy (PFM). Figure 2(f) shows two distinct PL signals on a $MoSe_2$-$WSe_2$ BL-HS on two different types of moiré superlattice. Type I (figure 2(f) upper panel) is the normal hexagonal moiré superlattice (see inset) whose PL consists of sharp peaks with uniform CP, consistent with earlier results[36,50]. Type II (figure 2(f) lower panel) is the distorted 1D moiré superlattice (see inset) due to strain, for which the corresponding PL consists of a single broad peak with minimal CP. Interestingly, the broad PL peak from type II moiré superlattice is strongly linearly polarized. The linear polarization direction tends to align with the direction of the primary structure of 1D moiré.

***Singlet and triplet interlayer excitons and other interpretations***

A number of optical studies on MoSe$_2$-WSe$_2$ BL-HS report the observation of singlet and triplet IELXs as adjacent peaks with opposite CP in PL measurements[49,53,54] (figure 2(g)). The energy separation between these two peaks is about 25-40 meV in agreement with the CB spin splitting of MoSe$_2$. Ciarrocchi et. al.[53], studying R-type BL-HS, assign the double interlayer peaks to singlet (lower energy) and triplet (higher energy) localized at the $R_h^X$ site of the moiré superlattice. For H-type BL-HS, the singlet (higher energy) and triplet (lower energy) are located at $H_h^h$ site[49,54]. The g-factors are different for IELXs in H-type (~15.2 for triplet, ~10.7 for singlet[49]) and in R-type (7.1±1.6 for triplet, -8.5±1.5 for singlet[53]) due to the contribution of spin, and different contribution of angular orbital momentum which depends on where the IELX is localized ($R_h^h, R_h^X, or\ H_h^h$)[55,56]. Singlet and triplet energy positions and intensities are highly tunable (more than 100 meV energy shift) similar to the moiré sharp peaks[50], either by out-of-plane electric field[53] or by electrical doping[49,53]. More interestingly, since electrical doping can change the relative intensity between singlet and triplet peaks, CP of the IELX emission can be switched from polarization-preserving to polarization-inverting regime[53].

Other studies also report the observation of the double IELX peaks with different interpretations. Hanbicki et. al.[57] show the emergence of double peaks with opposite CP, and assign the peaks to double indirect IELX in momentum space in agreement with density functional theory (DFT) calculations. Miller et. al.[58] assign the higher (lower) energy peak to momentum direct (indirect) IELX. On the other hand, Calman et. al.[59] interpret the lower (higher) energy peak as interlayer trion (exciton). Because of these different interpretations, additional experimental techniques and further studies are warranted.

*Interlayer exciton diffusion*

IELX diffusion represents a unique transport problem involving energy, charge and spins in TMD BL-HS[28,60-62]. Rivera et. al. have earlier demonstrated that valley-dependent exchange interaction between IELXs of MoSe$_2$-WSe$_2$ BL-HS in the same valley (K) can cause different repulsive forces for majority-valley (K) and minority valley (K') populations, resulting in valley-asymmetric drift and diffusion of IELXs[28]. The long-lived nature of IELXs in hBN-encapsulated $MoS_2$-$WSe_2$ and $MoSe_2$-$WSe_2$ BL-HS leads to diffusion on the order of a few micrometers, with dynamics manipulation by excitation power or electrical controls[60-62]. These previous diffusion studies suggest the free-diffusion picture for IELXs in absence of confinement. The absence of confinement could possibly be due to larger twist-angles or large lattice mismatch resulting in weaker moiré potentials, or from inhomogeneously averaged excitonic response.

Due to the presence of moiré potential, IELX diffusion in incommensurate BL-HS is expected to possess unique features distinct from commensurate BL-HS. While two different views for the IELXs as free or localized carriers exist, recent studies suggest that the presence of moiré potential has a strong influence on IELX diffusion in TMD BL-HS[45,61,63,64]. The moiré potential exhibits periodic energy modulation on the order of ~ 150 meV, deep enough to completely localize IELXs within the potential or impede their lateral diffusion.

The IELX diffusion in $MoSe_2$-$WSe_2$ BL-HS with commensurate stacking and corresponding absence of moiré potential, shows the strikingly long diffusion length on the order of a few micrometers, in contrast to the small diffusion lengths in the incommensurate BL-HS and presence of moiré potential (figure 3(a,b))[63]. Moreover, using spatially resolved transient absorption technique, it was observed that the IELX diffusion in $WS_2$-$WSe_2$ CVD-grown BL-HS with R-type stacking is suppressed compared to H-type stacking, due to the deeper moiré potential depth (figure 3(c,d))[45]. To observe the IELX dynamics, $WSe_2$ ($WS_2$) IALX resonance is pumped (probed), measuring the electron population residing in $WS_2$ layer after the interlayer charge transfer. In mechanically stacked $MoSe_2$-$WSe_2$ BL-HS with near 0° stacking, at low optical excitation density, the diffusivity of IELXs can be drastically reduced due to the effective trapping of IELXs by moiré potential. All recent diffusion studies provide a different perspective of spatial diffusion behavior of IELXs in the TMD BL-HS and suggest that the IELX diffusion is dominated by the competition between many-body interactions and localization by moiré potential.

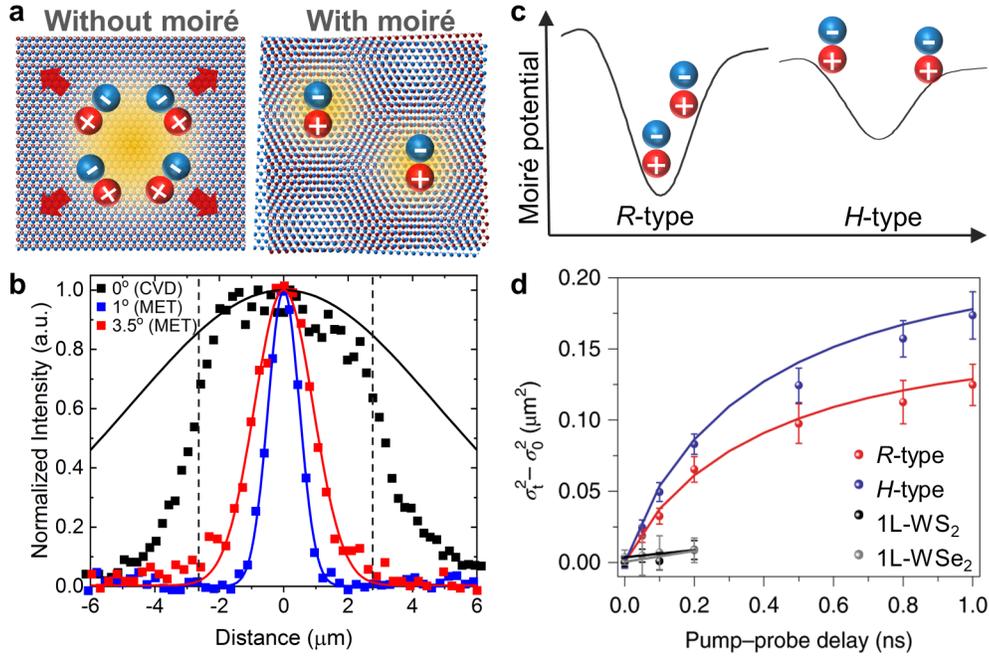

**Figure 3**. IELX diffusion influenced by moiré potential in TMD BL-HS. **a,** Schematic of IELX diffusion in the absence or presence of moiré potential in $MoSe_2$-$WSe_2$ BL-HS. **b,** Twist-angle dependent spatial profile of IELX diffusion at low temperature. Dashed lines indicate the boundary of BL-HS region in 0º sample. CVD and MET refer to samples prepared by chemical vapor deposition and mechanically exfoliation & transfer, respectively. **c,** Illustration of spatial landscape of moiré potential with different depth in R- and H-type stackings. **d,** IELX diffusion in the R-type and H-type $WS_2$-$WSe_2$ BL-HS are compared to the IALX diffusion in $WS_2$ and $WSe_2$ MLs. The plots show increasing mean squared distances travelled by IELXs with a non-linear dependence on delay time. $WSe_2$ ($WS_2$) IALX resonance is pumped (probed), measuring the electron population residing in $WS_2$ layer after the interlayer charge transfer. Panels reprinted and adapted with permission, **a-b** from ref. [63] (American Association for Advancement of Science), **d** from ref. [45] (Springer Nature).

**Intralayer excitons**

Properties of IALXs in a BL-HS can be modified due to the periodic moiré potential operating in the weak or strong-coupling regime. This moiré potential can mix momentum states separated by moiré reciprocal vectors, leading to additional absorption peaks with predicted topological properties, flat exciton dispersion and protection against disorder[11,19,37]. Jin et. al. tracked optical reflectance of a WSe$_2$-WS$_2$ BL-HS with carrier concentration, and measured three WSe$_2$ intralayer MX peaks (figure 4(a))[19]. All peaks blue-shift with increasing electron concentration but with varying levels of shift, related to different localization with respect to the moiré superlattice. The peaks vanish with increasing twist-angles signifying the importance of moiré superlattice (inset of figure 4(a)). The twist dependence is in qualitative agreement with calculations performed on a WS$_2$-MoS$_2$ BL-HS, considering moiré potentials without any domain formation (figure 4(b))[37].

The depth of moiré potential for intralayer MXs is relatively small compared to that for interlayer MXs[11]. Similar CP of intralayer MoSe$_2$ MXs (and trions) on a MoSe$_2$-MoS$_2$ BL-HS and a bare MoSe$_2$ ML has been observed (figure 4(c))[65]. This work also showed the evolution of interlayer exciton and trion peaks to higher and lower energy excitons ($X_H, X_L$) and trions ($T_H, T_L$) peaks in the BL-HS. These results suggest unchanged selection rules for IALX (with respect to the IALXs in bare ML) when modulated by the moiré potential, and preservation of rotational symmetry of transitions[11]. Further, the same work[65] measured a small change in MoSe$_2$ trion decay-time from 79 ps in bare ML, to 64 ps in the BL-HS for $T_H$ (72 ps for $T_L$). On the other hand, exciton dynamics are measurement-resolution limited and remain at ~14 ps. However, the modulation depth for intralayer MXs in a MoSe$_2$-MoS$_2$ BL-HS is ~ 10 meV,[65] whereas for a WSe$_2$-WS$_2$ BL-HS is ~ 30 meV[19]. Thus, in the case of WSe$_2$-WS$_2$ BL-HS, varied dynamics may be possible compared to weakly modulating moiré potentials.

The origin of these additional MX peaks is still under vigorous debate. The different additional peaks for the intralayer MX are considered to arise from localization at different high symmetry points in the real-space moiré superlattice. The evidence for these is indirect at best. For example, peak shift of intralayer MX peaks (for WSe$_2$ exciton in a WSe$_2$-WS$_2$ BL-HS) with doping has been attributed to localization of MX peaks, as well as electrons in WS$_2$ layer, with respect to the moiré superlattice[19]. On the other hand, reconstructed domain related intralayer MX bands have been invoked in a twisted WSe$_2$ homo BL-HS[66]. The field would benefit with direct nanoscale characterization of structure and optical properties[67].

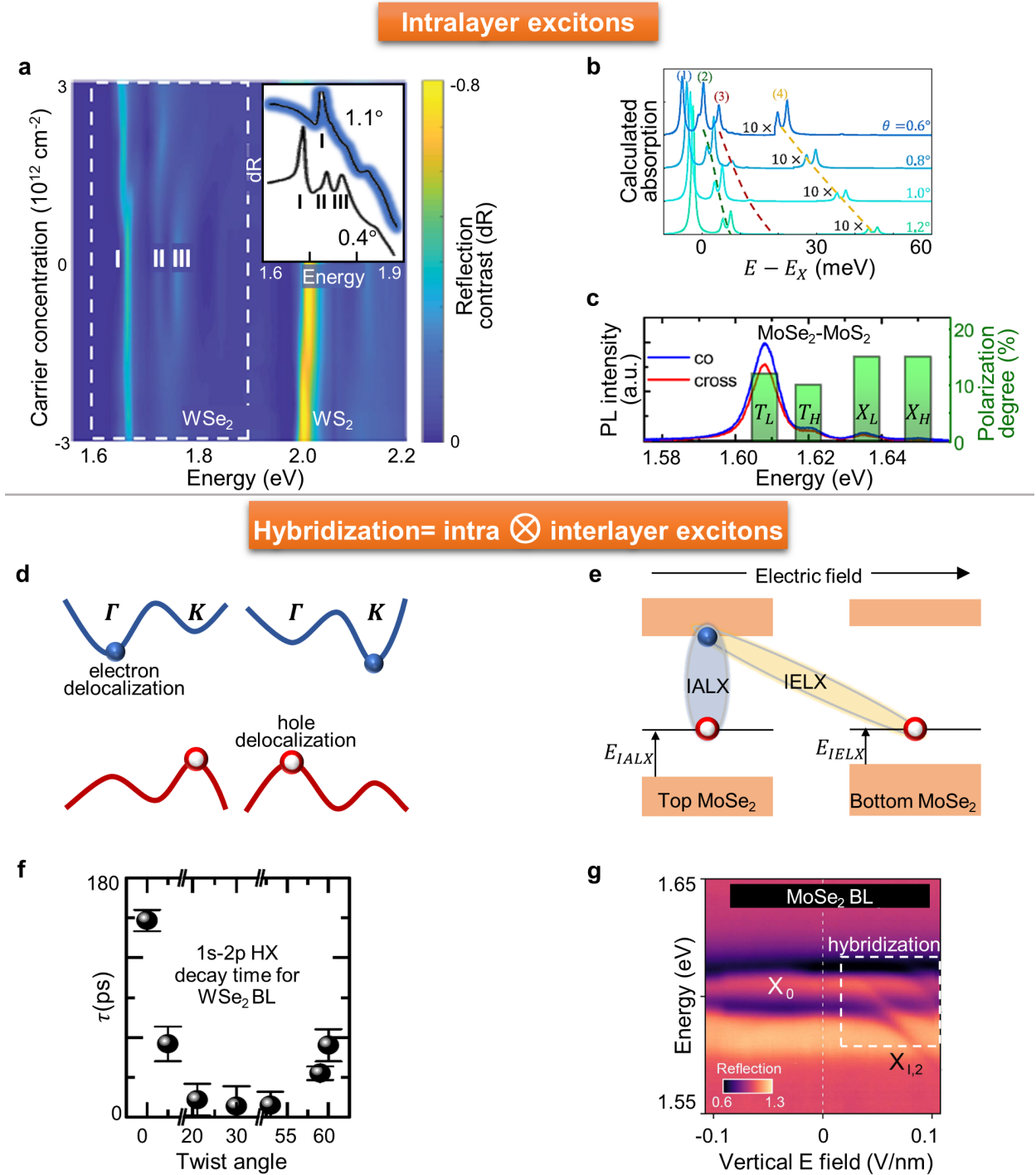

Figure 4: a-d, Intralayer excitons (IALX) modified by moiré potential. a, evolution of intralayer WSe$_2$ MXs in a WSe$_2$-WS$_2$ BL-HS with doping. Resonances II, III are additional intralayer MX bands, that disappear with larger twist-angle as shown in inset. b, Calculation of optical response in WS$_2$-MoS$_2$ BL-HS with different twist-angles. c, Valley polarization of intralayer MoSe$_2$ MXs (and trion) in MoSe$_2$-MoS$_2$ BL-HS. The intralayer exciton and trion

peaks evolve to higher and lower energy excitons ($X_H, X_L$) and trions ($T_H, T_L$) in the BL-HS. **d-g, Hybridized excitons (HX) d,** Schematic of delocalization of electrons or holes, depending on nature of band orbitals and relative band-ordering. **e,** Formation of HX in MoSe$_2$ homo BL-HS, where a vertical electric field tunes IELX to near-degeneracy with IALX, while sharing a hole state. **f**, Twist-angle variation of dynamics of Lyman-like 1s-2p exciton transitions in WSe$_2$ homo BL-HS. **g,** Electric-field dependent reflectance spectra of MoSe$_2$ homo BL-HS indicating Stark shift of HX (appearing as $X_{I,2}$) and anti-crossing (signature of hybridization) with IALX ($X_0$). Panels reprinted and adapted with permission, **a** from ref[19] (Springer Nature), **b** from ref[37] (American Physical Society), **c** from ref[65] (American Chemical Society), **f** from ref[68] (Springer Nature), **g** from ref[69] (Springer Nature).

**Hybridized excitons**

Hybridized excitons (HXs) are created when two conditions are met in a BL-HS: IELX formation due to type-II band alignment, and CB or VB degeneracy between the two layers[41,68-71]. HXs result from transfer of optical oscillator strength from IALXs to nearly degenerate IELXs and sharing of a common electron or hole state[71]. HXs inherit electric dipole from IELXs and optical dipole from IALXs, and thus, tunable by both electrical and optical fields.

Depending on nature of band orbitals and relative band-ordering, either electrons or holes are delocalized in the two layers (figure 4(d)). For example, in twisted MoSe$_2$ homo BL-HS (for 0° stacking), the CB minimum (VB maximum) is located at the $K$ ($\Gamma$) momentum-point. Since the orbitals contributing to $\Gamma$ are more delocalized (compared to $K$ point), the hole is delocalized over the two MLs, whereas the electron is constrained to one of the layers[69]. On the other hand, for twisted WSe$_2$ homo BL-HS, the electron is delocalized due to CB minima at $\Gamma$ point[68]. Direct momentum (K-K) IELX were also recently measured in MoS$_2$ BL-HS, with hybridization controllable by stacking order[71-73]. The picture is slightly more complicated for a hetero BL-HS, where disparate orbital nature of the two component layers must be taken in account.

In a MoSe$_2$-WS$_2$ BL-HS, CB is nearly degenerate for the two layers, and electron state is shared. Further, the momentum mismatch ($\Delta k$, due to twist) is surmountable due to moiré brillouin-zone enabled momentum conservation. The mixing of intralayer moiré bands with IELX states result in enhanced interlayer tunneling and formation of optically-bright HX. This mixing also results in a twist-angle dependence of HX[74], and for Lyman-like (level transitions in an atom) 1s-2p HX transitions (in twisted WSe$_2$ homo BL-HS[68]).

Due to finite electric dipole, Stark shift of HXs is far stronger than IALXs. In a dual-gated (top and bottom) configuration, both Stark shift of HX and avoided crossing have been observed (MoSe$_2$ homo BL-HS[69], MoSe$_2$-hBN-MoSe$_2$ [41]). Leisgang et. al. observed a large Stark splitting of 120 meV between two degenerate IELX and avoided crossing with the B-type IALX[73]. Avoided crossing is a characteristic feature of hybridization of resonances, and results from near-degeneracy of electric-field-tuned IELX and nearly unaffected IALX (figure 4(e,g))[69].

Twist-angle can tune the extent of interlayer hybridization, which results in varying temporal dynamics. For 1s-2p HX transitions in twisted WSe$_2$ homo BL-HS, lifetime decreases from 0° to 30° twist-angle followed by an increase till 60° (figure 4(f))[68]. By increasing the twist-angle beyond 0° (up-to 30°), electron-hole overlap in the same layer increases, leading to reduction in lifetime. Thus, at small (large) twist-angles HX have more interlayer (intralayer) nature and have long (short) lifetimes. With larger twist-angles, interlayer hopping becomes even weaker, which further reduces the exciton lifetime.

**Strongly correlated electronic states**

Following the exciting discoveries in twisted graphene homo BL-HS[23,75], TMD BL-HS have also been found to possess correlated quantum phenomena revealed by optical spectroscopy and transport measurements[39-41,76,77]. Recent theoretical work has shown that the moiré superlattice in TMD BL-HS can be used to simulate the 2D triangular spin-lattice Hubbard model[38]. The model with parameters such as band filling, hopping ($t$), and on-site repulsive interaction ($U$), can predict magnetization and correlations. In a TMD moiré superlattice, the realization of a single moiré band with two valley degrees of freedom, playing the role of spins, is useful to simulate this model system. For example, $U$ decreases as the moiré periodicity increases, while *t* can be modified with choice of component MLs and/or twist angle. Here, the half-filled moiré band can satisfy the condition for strongly correlated states, as the ratio of *U* to electron bandwidth substantially increases.

Optical spectroscopy coupled with electrical control has been used to investigate strongly correlated electronic phases by probing excitonic resonances. Regan et. al. implement optically-detected resistance and capacitance to measure the carrier concentration in a $WSe_2$-$WS_2$ BL-HS, revealing various doping-dependent insulating states[40]. The induced change in optical contrast (ΔOC) of the $WSe_2$ IALX resonance, under the electrical modulation of the sample's top gate, is directly proportional to the density of states of the moiré superlattice. The ΔOC shows strong gap-like features at n = $n_0$/6, $n_0$/3, and $n_0$/2 hole doping densities, where $n_0$/2 is defined as the density at which one hole occupies one moiré supercell (figure 5(a)). Such gap-like features are enhanced at higher electrical modulation frequency (30 kHz). These insulating states at 1/6 and 1/3 filling of the superlattice are assigned to generalized Wigner crystallization, while Mott insulating state[78,79] is revealed at one hole per superlattice, arising from Coulomb interaction between holes localized in same moiré superlattice (figure 5(b)). These correlated insulating states are absent in $WSe_2$-$WS_2$ BL-HS with a large twist-angle. Here, we note that the half-filling factor is defined as n=1 in this work[40], while $n_0$=1/2 is generally used in other works (and adopted in this review paper).

The Hubbard model also predicts different magnetic phases depending on the filling and controlling the interaction regime ($U/t$)[38,80-82]. Indeed, a Mott insulating state with antiferromagnetic (aFM) Curie-Weiss behavior was observed at half-filling of $WSe_2$-$WS_2$ moiré superlattice band, while possible signatures of phase-transition from an aFM to a weak ferromagnetic (FM) state were shown above half-filling (figure 5(c))[39]. Here, the magnetic moments of localized holes are probed using magneto-optical spectroscopy by measuring Zeeman splitting of MX resonances and corresponding g-factors. Particularly, a giant g-factor (~270) is observed between 0.5 and 0.7 fillings, while a g-factor of -4 (corresponding to IALX of a pristine ML)[55,56,83-85] is observed near zero and full fillings. The sensitivity of MXs on magnetic interactions is modeled by a molecular field $\lambda_X M$ ($\lambda_X$ = coupling constant; $M$ = sample magnetization). The analysis of temperature-dependent magnetic susceptibility $\chi$ suggests Curie-Weiss behavior with a negative Weiss-constant (Θ), thus indicating an aFM state in the high $U/t$ regime. Moreover, diverging $\chi$ and vanishing Θ near 0.6 filling factor (vertical dashed line in figure 5(c)) represents the possible existence of quantum phase transition from aFM to weak FM state, while the transition temperature was not accessible in this study.

In the context of strongly correlated phases in the Hubbard model, the twisted MoSe$_2$ BL-HS system with insertion of a single layer hBN (between the MoSe$_2$ MLs) has been introduced as a unique regime with long moiré periodicity and suppressed moiré potential[41]. Figure 5(d) shows the vertical external electric field ($V_E$) dependence of the differential reflectance spectrum (ΔR/R$_0$) at half-filling of the lowest moiré band. As the electric field increases, an abrupt shift of IALX resonance as well as the complete transfer of oscillator strength between polarons in top and bottom layers ($AP_{top}$ or $AP_{bot}$) is observed, indicating the complete transfer of all electrons from one MoSe$_2$ ML to the other. This abrupt transfer is indication of an interaction-induced incompressible Mott state in the lowest moiré band. Here the exciton-polaron is a many-body hybrid light matter state, and acts as a spectroscopic tool to measure interactions of excitons and electrons/holes[41].

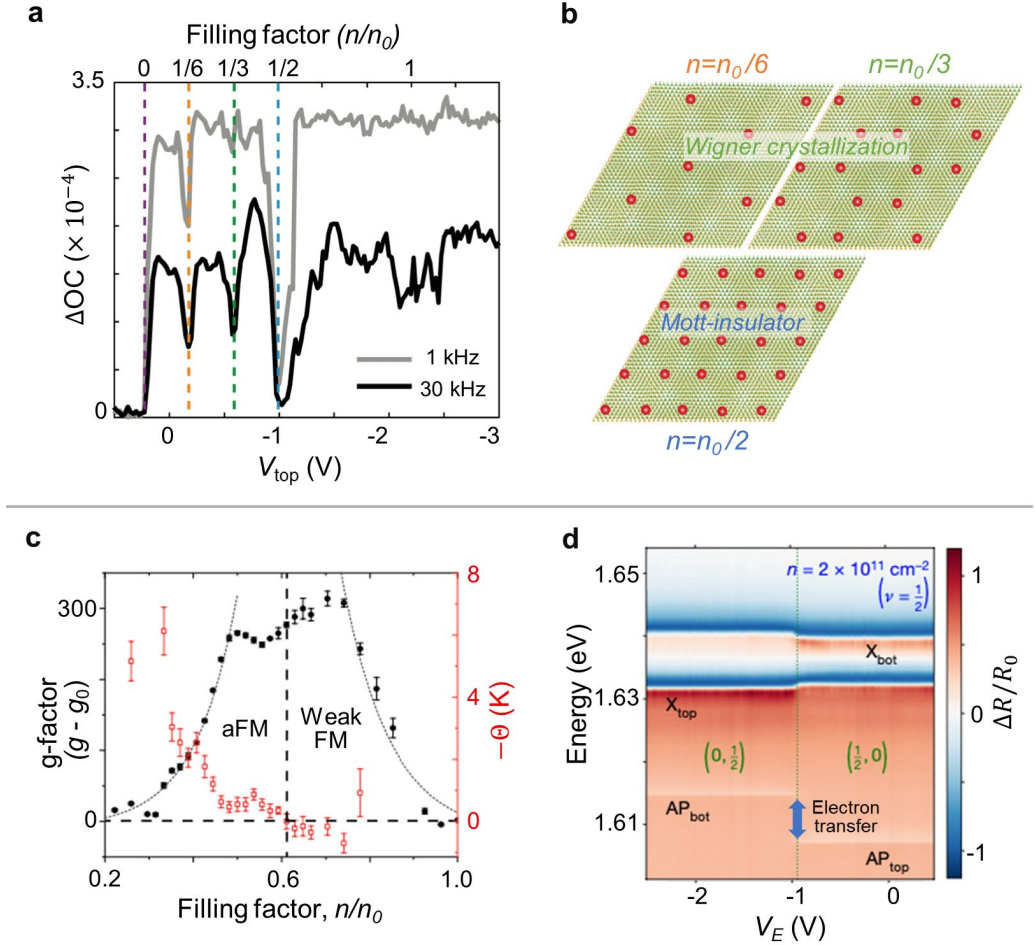

**Figure 5**. Strongly correlated electronic systems in TMD moiré superlattice. **a,** Doping dependent change of optical contrast (ΔOC) at 1 and 30 kHz shows strong gap-like features at n=$n_0$/6 (orange), $n_0$/3 (green), and $n_0$/2 (blue) hole doping levels in WSe$_2$-WS$_2$ BL-HS device. **b,** Description of Wigner crystal (n=$n_0$/6, $n_0$/3) and Mott insulator states (n=$n_0$/2) in a WSe$_2$-WS$_2$ moiré superlattice. **c,** Filling factor dependence (at 1.65 K) of g-factor and Weiss constant Θ. Sign inversion of Θ at filling factor ~ 0.6 suggests transition from anti-ferromagnetic (aFM) to weak ferromagnetic (FM) state. **d,** Electric field dependent mapping of differential reflectance at half filling ($v = 1/2$). $X_{top}$ and $AP_{top}$ ($X_{bot}$ and $AP_{bot}$) refer to IALX and attractive polaron in top (bottom) layer, respectively. At $V_E$~ -1V, an abrupt transfer of electrons from one ML to the other takes place. Panels reprinted and adapted with permission, **a,b** are adapted from ref[40] (Springer Nature), **c** from ref[39] (Springer Nature), and **d** from ref[41] (Springer Nature).

**Mapping of moiré potential and reconstruction of domains**

So far, we have discussed the effect of moiré potential on opto-electronic properties, without providing direct evidence of moiré superlattice in real space. In this section, we provide real-space nanoscale mapping of rigid moiré superlattices and possibilities of domain reconstruction measured by various techniques including transmission electron microscope (TEM)[19,69,86], scanning electron microscopy (SEM)[66,87], conducting-atomic force microscopy (c-AFM)[21], scanning tunneling microscope (STM)[42,88] and PFM[48,89]. In this section, for consistency of nomenclature with the work in this field, we define $R_h^h, R_h^X, R_h^M$ as AA, AB and BA sites respectively (for R, parallel stacking). For H stacking (anti-parallel), we define $H_h^h, H_h^X, H_h^M$ as ABBA, AA and BB sites respectively.

For a WSe$_2$-WS$_2$ BL-HS (R stacking, twist-angle > 0.3°), moiré unit cell of length 8 nm is observed using annular dark-field scanning TEM (ADF-STEM) (figure 6(a)), consistent with the rigid layer moiré superlattice picture[19]. Earlier high-resolution TEM (HRTEM) work also showed the presence of moiré patterns in twisted-MoS$_2$ layers.[90] STM provides complementary high-resolution spatial mapping of density of states, which in-turn are affected by layer-layer interactions.[91] STM mapping of WSe$_2$ homo BL-HS (twist-angle ~ 3°) shows a moiré period of ~ 6.3 nm with visible contrast for AA, AB and BA sites (figure 6(b))[88]. Similar STM work on CVD grown MoS$_2$-WSe$_2$ BL-HS (conformal growth, no twist-angle)[42,92] shows moiré superlattice period of 8.7 nm (figure 6(c))[92]. In these samples, only the moiré superlattice is observed with no reconstruction (which we explain below).

Structural relaxation of moiré superlattice, at the level of DFT, force-fields or multi-scale modelling of interlayer binding energy densities, reveal that AB (or BA) triangular stacking-domains are energetically favorable compared to AA domains for R stacking (figure 6(d))[21,86,93]. On the other hand, ABBA domains are energetically favorable for H stacking[94,95]. Thus, energy of the system can be minimized through lattice reconstruction, via expansion of energetically-favorable stacking domains and creation of domain walls. The level of atomic reconstruction is related to two competing parameters: interlayer stacking energy, which drives atomic reconstruction; and intralayer strain, which opposes reconstruction. In simpler terms, interlayer stacking energy drives domain formation (and stretching of layers to form commensurate structure), whereas the intralayer bonding aims to maintain pristine, rigid lattice. Below a critical twist-angle ($\theta_c$), potential energy gradient due to interlayer stacking energy can dominate the restoring force due to intralayer strain, and can result in reconstruction[89].

For R stacked BL-HS with small twist-angles, significant lattice reconstruction has been observed ($WS_2$-$MoS_2$ BL-HS, 1.5° twist-angle) by using ADF-STEM[86]. Reconstructed domains in $WSe_2$ homo BL-HS were measured using PFM (figure 6(e)), where contrast results from modulation of piezoelectric tensor by moiré superlattice potential or reconstructed domains[48,89]. Further, inversion symmetry of AB and BA domains can be broken by explicit tilting of the sample in DF-TEM, thus enabling direct domain visualization in twisted $MoSe_2$ (~ 0°) homo BL-HS (figure 6(f))[69]. DF-TEM uses an aperture to filter a particular Bragg diffracted peak, thus providing a direct mapping of stacking order[96,97]. Similarly, tilted-SEM has measured varying sizes of domains in $WSe_2$ homo BL-HS (~ 0°) while differentiating between AB and BA domains, and is an encouraging new technique due to its compatibility with hBN encapsulated BL-HS prepared for optical experiments[66,87]. For increasing twist-angle, it has been observed (via TEM) that triangular domain contrast is weakened and becomes close to one-directional fringes, a characteristic of simple rotational moiré structure without reconstruction[21,96].

For H stacking, distorted hexagonal domains (ABBA) are energetically-favorable (figure 6(g))[86,93,95], but $\theta_c$ is much smaller compared to the case of R stacking. For $MoS_2$ homo BL-HS (H stacking, 0.25°), distorted hexagonal domains (with varying size) were observed using ADF-STEM (figure 6(h))[86]. Similarly, c-AFM measured reconstruction in $WSe_2$-$MoSe_2$ BL-HS (H stacking, ~ 0°) (figure 6(i))[21].

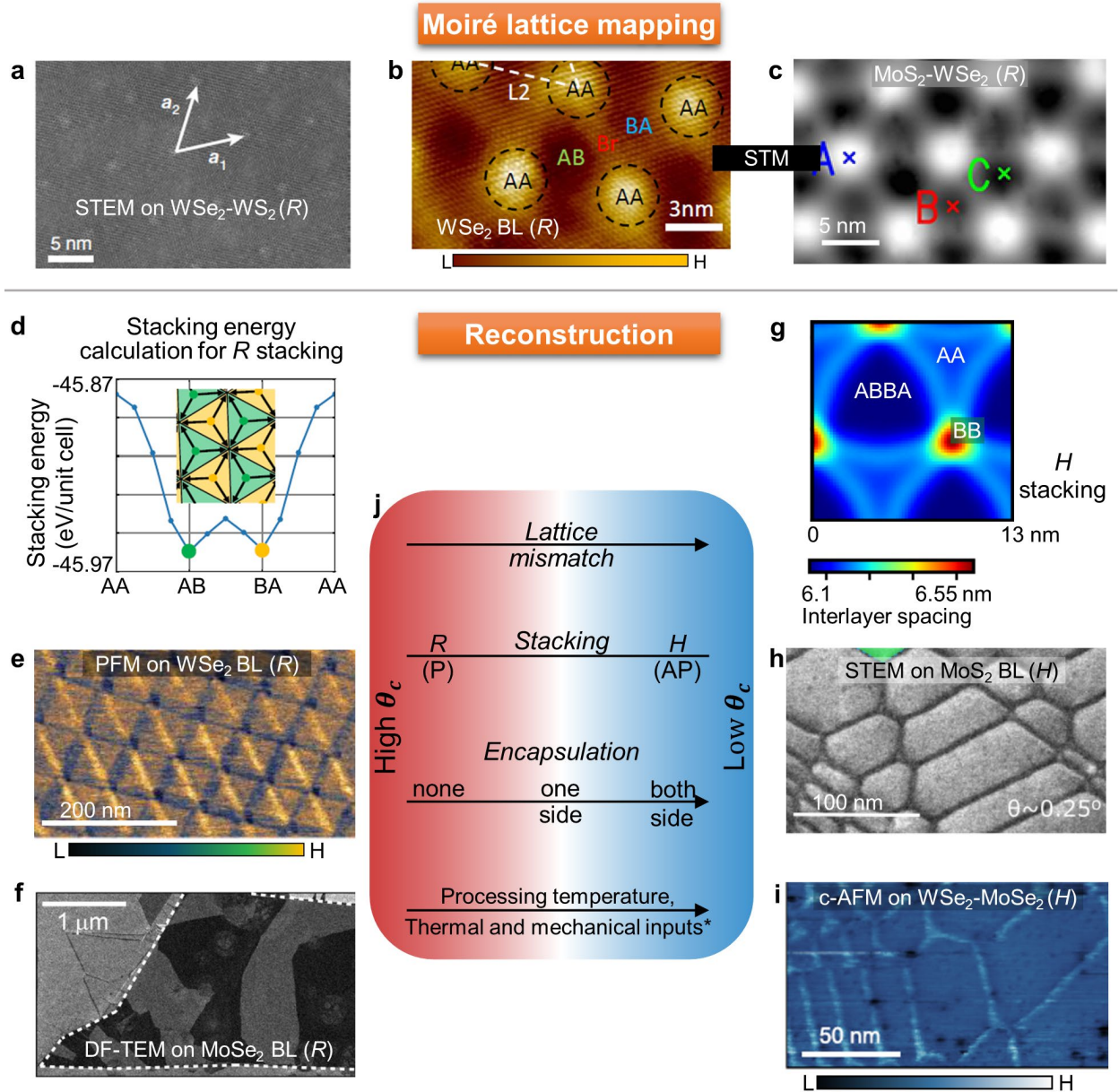

**Figure 6**: **Rigid moiré superlattice (a-c)**. **a,** WSe$_2$-WS$_2$ BL-HS (annular dark-field scanning transmission electron microscopy (ADF-STEM), R-type stacking, ~0° twist-angle); **b,** WSe$_2$ homo BL-HS (scanning tunneling microscopy (STM), R, ~3°). **c,** CVD grown MoS$_2$-WSe$_2$ BL-HS (STM, R, 0°). **Reconstruction and domains (d-j).** **d,** Calculation of stacking energy for energetically favorable AB and BA domains for R stacking. Reconstructed domains observed in **e,** WSe$_2$ homo BL-HS (piezo-force microscopy (PFM), R, ~0°); **f,** MoSe$_2$ homo BL-HS (tilted-dark-field TEM, R, ~0°). **g,** Calculated reconstructed domains for H stacking. Reconstructed distorted hexagonal domains observed in **h,** MoS$_2$ homo BL-HS (ADF-STEM, H stacking, 0.25°); **i,** WSe$_2$-MoSe$_2$ BL-HS (conducting

atomic force microscopy (c-AFM), H, ~0°) **j,** summary of variation of critical twist-angle ($\theta_c$) with various HS component and processing relevant properties. Panels reprinted and adapted with permission, **a** from ref[19] (Springer Nature), **b** from ref[88] (Springer Nature), **c** from ref[92] (American Chemical Society), **d,i** from ref[21] (American Chemical Society), **e** from ref[89] (Springer Nature), **f** from ref[69], **g** from ref[95] (American Physical Society), **h** from ref[86] (Springer Nature).

It is important to remember that the calculations shown in figure 6(d,g) are thermodynamic equilibrium calculations. Driving forces are needed to ascend possible thermodynamic barriers and force the system towards equilibrium, even for twist-angles < $\theta_C$. Possible driving forces are thermal annealing process or mechanical inputs (for example, shear) usually present during HS preparation. On the other hand, hBN encapsulation can change the amount of reconstruction by providing opposing driving forces, as shown explicitly for the case of graphene-hBN[91], and hinted at for WSe$_2$ homo BL-HS[88]. For materials with similar unit cell size (MoSe$_2$-WSe$_2$, or homo-BLs), $\theta_c$ is much larger compared to materials with significant lattice mismatch (example MoS$_2$-WSe$_2$). We summarize some of these ideas in the figure 6(j). Thus, for low lattice mismatch, R stacking, no encapsulation, or high processing temperature, $\theta_C$ should be large.

**Conclusion and Outlook**

In this review paper, we have reviewed the state of the twistronics field focusing on TMD BL-HS. We have discussed moiré IELXs and IALXs and their hybridization, effect of moiré superlattice on diffusion, strongly correlated states, and moiré superlattice mapping. Tremendous progress has been made in the past few years, which has opened many avenues of research as well as raised fundamental questions about origin of spectral features.

Moiré IELXs promise to be new sources of 2D-array of identical quantum emitters, with applications in optoelectronic devices, and quantum communication. These quantum emitters are highly tunable by various experimental knobs, including vertical electric field[50,53], doping[53], and strain[48]. Moiré IELX dipole can point in or out-of-plane, depending on whether exciton is of spin-singlet or triplet nature, and on exciton location within the moiré superlattice [24]. IELX diffusion is also affected by the moiré potential, with different diffusion regimes for R and H-type stacking. An IELX dipole is thus tunable for energy, polarization and strength using existing opto-electronic means, and thus relevant for opto-electronic applications.

IALXs are also modified by the moiré potential, with the twist-angle directly tuning the energy spacing and oscillator strength of additional resonances. An outstanding question is how the dynamics of IALXs are affected by the strength of the moiré potential. Similar questions arise for HXs, with the additional complication of momentum space band-ordering. Since HXs are tunable by both optical and electrical fields, they offer promise in integrated photonics as active materials[4], as well as in fundamental measurements of cavity-coupling dynamics. Topological properties and presence of flat bands (in twisted homo BL-HS) are also exciting fields of research and we encourage extensive theoretical and experimental work.

TMD BL-HS represents a promising solid-state platform to simulate triangular-lattice Hubbard models based on moiré superlattice. Feature-rich strongly correlated physics can be explored by combination of optical spectroscopy and electrical measurements for TMD BL-HS. Moiré potential depth in different BL-HS offers modulation of Hubbard model tuning parameters and corresponding modulation of correlations. The physics of moiré-systems with low filling factors remains to be explored, requiring work to prepare low resistance contacts and low carrier concentration samples.

For a meaningful comparison of theory and experiment regarding MXs, direct nanoscale optical mapping techniques are sorely needed. With large spatial resolution (~ μm) of far-field optics, evidence for MXs are often reported with conflicting data and interpretations. Near-field optical techniques such as cathodoluminescence[62] and tip-enhanced photoluminescence[98-100] offer the possibility of mapping of individual moiré supercells with spatial resolution of a few nanometers. Cathodoluminescence in a TEM has the added advantage of direct optical-structural correspondence, and simultaneous measurement of stacking order. Such techniques can naturally lead to measurement of effects of strain and atomic scale defects.

Mapping techniques (PFM, TEM, STEM, and STM) have established the presence of rigid moiré superlattice in BL-HS, with evolution to reconstructed domains for small twist-angles and low lattice mismatch. Since evidence of reconstruction is growing, the field now needs to look at the processing conditions more carefully with systematic studies critically needed. A concern is the varying size of reconstructed domains across micrometer length scales, most probably resulting from pinning at defect sites or imperfections, which naturally lead to spatially varying opto-electronic properties. Further, how reconstruction changes the optical properties of TMD BL-HS and whether novel physical states can be created, are outstanding topics for investigation[86]. Interesting avenues of research will be as to how the moiré picture is modified in reconstructed domains.

**Acknowledgements**

A.S. would like to acknowledge funding from Indian Institute of Science start-up grant. K.T acknowledges support from the National Research Council postdoctoral research associateship. J. C. would like to acknowledge funding from the National Science Foundation via grant DMR-1808042 and the U.S. Department of Energy, Basic Energy Science program via grant DE-SC0019398. The authors would like to thank for suggestions by Hsun-Jen Chuang, Paulo Eduardo de Faria Junior, Tomasz Woźniak, Prabal Maiti, Manish Jain and Xiaoqin Li.

**Author Contributions**

All authors contributed to writing of the review paper.

**Competing Interests statement**